\documentclass[aps,prb,twocolumn,showpacs,amsmath,superscriptaddress,10pt]{revtex4-1}
\usepackage{amssymb}
\usepackage{amsmath}
\usepackage{graphicx}
\usepackage{float}
\usepackage{ulem}
\usepackage{bm}
\usepackage{dcolumn}
\usepackage{mathrsfs}
\usepackage{subfigure}
\usepackage{braket}
\usepackage{graphicx}
\usepackage{braket}
\usepackage{color}
\usepackage{bbold}

\usepackage{enumitem}
\newcommand{\cosinv}{\cos^{-1}}
\def\blue{\textcolor{blue}}
\def\red{\textcolor{black}}
\def\bea{\begin{eqnarray}}
\def\eea{\end{eqnarray}}
\def\bal{\begin{aligned}}
\def\eal{\end{aligned}}

\begin{document}
\title{Counter-propagating edge states in Floquet topological insulating phases}

\author{Muhammad Umer}
\affiliation{Department of Physics, National University of Singapore, Singapore 117551, Republic of Singapore}
\author{Raditya Weda Bomantara}
\email{Raditya.Bomantara@sydney.edu.au}
\affiliation{Department of Physics, National University of Singapore, Singapore 117551, Republic of Singapore}
\affiliation{Centre for Engineered Quantum Systems, School of Physics, University of Sydney, Sydney, New South Wales 2006, Australia}
\author{Jiangbin Gong}
\email{phygj@nus.edu.sg}
\affiliation{Department of Physics, National University of Singapore, Singapore 117551, Republic of Singapore}

\begin{abstract}
Nonequilibrium Floquet topological phases due to periodic driving  are known to exhibit rich and interesting features with no static analogs. Various known topological invariants usually proposed to characterize static topological systems often fail to fully characterize Floquet topological phases.  This fact has motivated extensive studies of Floquet topological phases to better understand nonequilibrium topological phases and to explore their possible applications. Here we present a theoretically simple Floquet topological insulating system that may possess an arbitrary number of counter-propagating chiral edge states.  Further investigation into our system reveals another related feature by tuning the same set of system parameters, namely, the emergence of almost flat (dispersionless) edge modes. \red{In particular, we employ two-terminal conductance and dynamical winding numbers to characterize counter-propagating chiral edge states.} We further demonstrate the robustness of such edge states against symmetry preserving disorder. Finally, we identify an emergent chiral symmetry at certain sub-regimes of the Brillouin zone that can explain the presence of almost flat edge modes. Our results have exposed more interesting possibilities in Floquet topological matter.

\end{abstract}

\maketitle

\section{Introduction}\label{sec_intro}
Following the discovery of quantum Hall effect \blue{\cite{Klitzing1980}}, topological explanation of quantized charge transport \blue{\cite{Thouless1982,Thouless1983,Kane2005,Kane2005a}} and experimental discovery of topological materials \blue{\cite{Bernevig2006,Koenig2007}}, topological phases of matter has emerged as one main research topic in condensed-matter physics. Here, topological phases of matter refer to systems which are protected not only by their underlying (internal and/or spatial) symmetry, but also by their inherent topological structure characterized by certain quantized invariants. These invariants, which are usually defined in the systems' bulk, manifest themselves as robust edge states at the systems' boundaries  \blue{\cite{Kane2005}}, thus leading to the so-called bulk-boundary correspondence \blue{\cite{Jackiw1976}}.

Depending on the specific system and symmetries under consideration \blue{\cite{Schnyder2008,Chiu2016,Roy2017}}, such edge states commonly take the form of either gapless chiral edge states in Chern insulators (superconductors) \blue{\cite{Hasan2010, Qi2011}} or flat (dispersionless) edge states in chiral/particle-hole symmetry protected topological insulators (superconductors) \blue{\cite{Su1979,Qi2008}}, Weyl semimetals \blue{\cite{Burkov2011,Hosur2013, Xu2015}}, and nodal line semimetals \blue{\cite{Burkov2011a,Mullen2015,Bian2016,Yu2015}}. Due to their topological origin, these edge states are protected against a wide range of symmetry preserving perturbations \blue{\cite{Kobayashi2013,Shapourian2016,Chen2019}}, which may lead to potential applications in designing robust electronic/spintronic devices as well as in fault-tolerant quantum information processing.

While originally topological phases are defined in terms of ground states of equilibrium systems, their studies in out of equilibrium systems, i.e., in the presence of time-periodic driving, have been extensively carried out since the last decade \blue{\cite{Oka2009,Kitagawa2010, Lindner2011,Derek2012,Zhou2018a, Asboth2013, Asboth2014, Rudner2013, Roy2016,
Lababidi2014,Zhou2014a,Ho2014,Zhou2014, Fulga2016,Xiong2016, Zhou2018}}. This leads to a variety of the so-called Floquet topological phases such as Floquet topological insulators \blue{\cite{ Asboth2014, Rudner2013, Roy2016,Lababidi2014,
Zhou2014a,Ho2014, Zhou2014, Fulga2016,Hockendorf2019,Xiong2016, Zhou2018}}, Floquet topological superconductors, \blue{\cite{Jiang2011,
Liu2013,Tong2013}}, Floquet Weyl semimetals \blue{\cite{Bomantara2016, Bomantara2016a}} and Floquet nodal line semimetals \blue{\cite{Wang2017,Li2018}}. Here, the term ``Floquet" refers to the application of Floquet theory \blue{\cite{Shirley1965,Sambe1973}} to characterize the topology of these time-periodic systems.

The time-dependence of the Hamiltonian describing Floquet systems indicates that energy is no longer conserved and is replaced by an analogous quantity termed quasienergy, defined only modulo the driving frequency $\omega$. Consequently, topological phase transitions under various symmetry considerations are not only due to band closing around zero quasienergy, but also due to that around $\omega/2$ quasienergy.  Because a Floquet eigenstate with quasienergy $\omega/2$ necessarily has an eigenphase $\pm\pi$, a quasi-energy gap at quasi-energy 0 and $\omega/2$ is also referred to as the zero or the $\pi$ (eigenphase) gap.   As such,  Floquet topological matter under the open boundary conditions may possess edge states in the zero or $\pi$ gap, or both.  Existing topological invariants defined for static topological systems may therefore not fully characterize the edge states of Floquet topological phases. Indeed, great efforts have been devoted in recent years to define new Floquet topological invariants. These include the dynamical winding number \blue{\cite{Rudner2013}}, scattering matrix invariants \blue{\cite{Fulga2016}}, and the symmetric time-frame winding number \blue{\cite{Asboth2013, Asboth2014}}, to name a few.

Floquet topological phases have attracted much attention not only due to the additional tunability offered by the periodic drive to switch between various distinct topological phases, but also due to their potential to exhibit remarkable features with no static counterpart. For example, certain Floquet topological insulators are known to exhibit nontrivial counter-propagating edge states (that is, with opposite chiralities), which arise when chiral edge states with opposite chirality localizes at the same boundary but around different quasimomenta \blue{\cite{Lababidi2014,Zhou2014a,Ho2014, Zhou2014, Fulga2016,Hockendorf2019}}. However, such counter-propagating edge states are not necessarily topologically protected. It is also worth noting that co-propagating edge states can respectively exist  at the zero and the $\pi$ gap.  This leads to the possibility that a nonequilibrium topological system has zero Floquet band Chern number but still hosts chiral edge states \blue{\cite{Rudner2013,Lababidi2014}}.  Another potentially important feature of Floquet topological phases is their capability to host a large number of co-propagating chiral edge states \blue{\cite{Zhou2014, Xiong2016, Zhou2018a}}, which are characterized by a large dynamical winding number \blue{\cite{Rudner2013}} and can in principle be probed via two-terminal conductance \blue{\cite{Fulga2016,Yap2017,Yap2018}}.

The purpose of this article is to present yet other interesting features of Floquet topological phases, using a theoretically simple two band model. First, we report the possibility of having an arbitrarily large number of robust counter-propagating edge states in both zero and $\pi$ quasienergy gaps. Unlike the arbitrarily large number of co-propagating chiral edge states observed in Ref.~\blue{\cite{ Zhou2018}}, the counter-propagating nature of the edge states observed here indicates that even the dynamical winding number of Ref.~\blue{\cite{ Rudner2013}} can no longer fully describe them. We thoroughly study the phase transitions associated with the emergence of more counter-propagating edge states as we tune the system parameters.  \red{Moreover, we discuss the topological behaviour of the system in terms of the dynamical winding number and two-terminal conductance to build the bulk-boundary correspondence for these edge states, thus clearly proposing a way to account for the number of pairs of these counter-propagating edge states.} Second, we briefly study the impact of symmetry-preserving disorder to verify the robustness of such counter-propagating  edge states. Finally, as a side result in the same system, we show that for certain values of system parameters, our model system may host almost flat edge states coexisting with other chiral edge states. These almost flat edge states are explained by the emergence of certain local chiral symmetry at many isolated points in the Brillouin zone away from  the rest of chiral edge states.

This article is structured in the following way. In section \blue{\ref{sec_model}}, we introduce our system and analyse its symmetries to determine its topological class. In section \blue{\ref{sec_CP}}, we show how the proposed system may host many counter-propagating edge states. We comprehensively analyse the system's Floquet operator to analytically locate the topological phase transitions in section \blue{\ref{subsec_CP_bandclosing}}. \red{In section \blue{\ref{subsec_CP_TI}}, we establish the bulk-boundary correspondence of the system in terms of its two-terminal conductance and dynamical winding number.} In section \blue{\ref{subsec_Disorder}}, we introduce symmetry-preserving disorder to our model system and verify the robustness of its edge states. In section \blue{\ref{sec_flat_edge}}, we show that in certain parameter regime, the system hosts almost flat (dispersionless) edge states at zero and/or $\pi$ eigenphase, explained by an emergent chiral symmetry for many isolated points in the Brillouin zone. Finally, we conclude this paper by summarizing our results in section \blue{\ref{sec_sum}}.

\section{Model description and symmetry analysis}\label{sec_model}
We consider a two dimensional lattice with nearest neighbour hopping and on-site mass term. Each site in the lattice consists of two sub-lattice degrees of freedom denoted by ${\bf \sigma}$. Time periodicity is introduced by employing a three-step quench protocol such that the system Hamiltonian in momentum representation is given as,
\begin{flushleft}
\bea
\bal
H({\bf k}, t) =
\begin{cases}
H_{1}({\bf k}) = 3\gamma_{1}\sigma_{x} &T \le t < T + \frac{T}{3}\\
H_{2}({\bf k}) = 3\gamma_{2}\sigma_{y}&T + \frac{T}{3} \le t < T + \frac{2T}{3}\\
H_{3}({\bf k}) = 3\gamma_{3}\sigma_{z}&T + \frac{2T}{3} \le t < 2T
\end{cases} \;,
~~~\label{EQ:Ham_Quench}
\eal
\eea
\end{flushleft}
where $T$ is the time period of the drive, $\gamma_{1} = J_{1}\sin(k_{x}),~ \gamma_{2} = J_{2}\sin(k_{y})$ and $\gamma_{3} = J_{3}[M + \cos(k_{x}) + \cos(k_{y})]$ are functions of two quasi-momenta $k_{x}$ and $k_{y}$. $J_{1}~(J_{2})$ is the hopping parameter in the $x~(y)$-direction during the first (second) step of the quench, and $J_{3}$ ($J_{3}M$) is the hopping parameter in both spatial dimensions (the on-site mass term) during the third step of the quench.

In passing, we would like to acknowledge that the above mentioned model has been previously studied in Ref.~\blue{\cite{ Zhou2018}}, which is based on the Floquet generalization of the Qi-Wu-Zhang (QWZ) model representing a minimal Chern insulating system \blue{\cite{Qi2006}}. Although this work uses the same model as in Ref. \blue{\cite{ Zhou2018}}, our work is different from that of Ref. \blue{\cite{ Zhou2018}}, which used the model to highlight the possibility of generating an arbitrary number of co-propagating chiral edge states. Here, we show that by appropriately tuning some other system parameters, two interesting features are observed. First, the same model is also capable of generating an arbitrary number of counter-propagating edge states, which cannot be captured by the dynamical winding number \blue{\cite{ Rudner2013}} alone. \red{As we will show below, an appropriate combination of dynamical winding number and the two-terminal conductance \blue{\cite{ Fulga2016}} can be constructed to fully characterize such counter-propagating edge states.} Second, at certain parameter values, almost flat (dispersionless) edge states emerge in the system and coexist with chiral and counter-propagating edge states, which leads to the emergence of chiral symmetry at some points in the Brillouin zone. To date, some features of Floquet topological phases (e.g., the presence of chiral edge states around zero and $\pi$ (eigenphase) gaps), have been experimentally demonstrated in graphene \blue{\cite{Mciver2020}}, photonic \blue{\cite{ Rec2013,Maczewsky2017,Mukherjee12017,Cheng2019}} and acoustic \blue{\cite{Peng2016}} systems. It is thus expected that a suitable modification of such experiments can be carried out to observe the results presented in this work.

The time-dependent Schr\"{o}dinger equation of the system is given as $i\hbar\frac{\partial}{\partial{t}}\mid\Psi_{n}({\bf k}, t)\rangle = H({\bf k},t)\mid\Psi_{n}({\bf k}, t)\rangle$, where $H({\bf k},t+T) = H({\bf k},t)$ and $T=\frac{2\pi}{\omega}$ being the period of the drive. Following the general practice in studies of Floquet topological phases, we now employ Floquet theory in quantum mechanics \blue{\cite{Shirley1965,Sambe1973}}. To this end, we first define a Floquet operator as the one-period time evolution operator, i.e., $\hat{\mathbb{T}}e^{-\frac{i}{\hbar}\int_{0}^{T}\hat{H}({\bf k}, t)dt}$, where $\hat{\mathbb{T}}$ is time ordering operator. Solving the Floquet eigenvalue equation $\hat{\mathbb{T}}e^{-\frac{i}{\hbar}\int_{0}^{T}\hat{H}({\bf k}, t)dt}\mid\Psi_{n}({\bf k})\rangle = e^{-i\Omega_{n}({\bf k})T/\hbar}\mid\Psi_{n}({\bf k})\rangle$ leads to a spectrum of eigenphases $\Omega_n({\bf k})$ termed quasienergies ($n$ being the band index), which replace the role of energies in such a time-periodic system. By construction, quasienergies are only defined modulo $\omega=\frac{2\pi}{T}$, which in this paper are taken to be within $\left(-\frac{\pi}{T},\frac{\pi}{T}\right]$. In this case, topological phase transitions occur when two quasienergy bands touch, while topological invariants are usually defined in terms of Floquet eigenstates $\mid\Psi_{n}({\bf k})\rangle$ when each quasienergy band is well separated from one another (away from the topological phase transition regime).

The Floquet operator associated with Eq.~(\blue{\ref{EQ:Ham_Quench}}) can be explicitly written as
\bea
\bal
U({\bf k}) = e^{\frac{-iH_{3}({\bf k})}{3}}e^{\frac{-iH_{2}({\bf k})}{3}}e^{\frac{-iH_{1}({\bf k})}{3}} \;,
~~\label{EQ:Floquet_Bulk}
\eal
\eea
where we have fixed $\hbar = T = 1$ for the rest of this paper (Hence, quasienergy is the same as Floquet eigenphase below).  Equation~(\blue{\ref{EQ:Floquet_Bulk}}) can be recast in the form,
\bea
U({\bf k}) = d_{0}\sigma_{0} - i(d_{x}\sigma_{x} + d_{y}\sigma_{y} + d_{z}\sigma_{z}) \;,
\label{EQ:Floquet}
\eea
where $\sigma_{0}$ is the $2\times 2$ identity matrix and ${\bf \sigma}$ are Pauli matrices representing the sub-lattice degrees of freedom. $d_{0}, d_{z}$ and $d_{x}, d_{y}$ are even and odd real functions  under $\left(k_{x},k_{y}\right) \rightarrow \left(-k_{x},-k_{y}\right)$, whose exact expressions are detailed in Appendix~\blue{\ref{App_bandclosing}}. The quasienergies are then given as $\Omega = \pm\cosinv(d_{0})$, where +(-) denotes the upper (lower) band. The Floquet operator can also be represented in terms of effective Hamiltonian such that $U({\bf k}) = e^{-iH_{eff}({\bf k})}$ where $H_{eff} = -i\log[U({\bf k})]$. Thus effective Hamiltonian from Eq. (\blue{\ref{EQ:Floquet}}) can be written as, $H_{eff}({\bf k}) = |\Omega|[\frac{d_{x}\sigma_{x} + d_{y}\sigma_{y} + d_{z}\sigma_{z}}{\sqrt{d^{2}_{x} + d^{2}_{y} + d^{2}_{z}}}]$, where $|\Omega| = \cosinv[d_{0}]$.

The topological classification of Floquet topological phases based on their internal symmetries are studied in Ref.~\blue{\cite{Roy2017}}. In particular, the charge-conjugation/particle-hole (PH $= \mathcal{PK}$), time reversal (TR $= \mathcal{TK}$) and chiral ($\mathcal{C}$) symmetry operations satisfy
\bea
\bal
\mathcal{P}U^{*}({\bf k})\mathcal{P}^{\dagger} &= U(-{\bf k})\\
\mathcal{T}U^{*}({\bf k})\mathcal{T}^{\dagger} &= U^{\dagger}(-{\bf k})\\
\mathcal{C}U({\bf k})\mathcal{C}^{\dagger} &= U^{\dagger}({\bf k}) \;,
\label{EQ:Floquet_Symm}
\eal
\eea
respectively, where $\mathcal{P}, \mathcal{T}$ and $\mathcal{C}$ are unitary operators and $\mathcal{K}$ is the complex conjugation operator. It can be verified that Eq.~(\blue{\ref{EQ:Floquet}}) respects the charge-conjugation symmetry $\mathcal{P} = \sigma_{x}$,
\bea
\bal\nonumber
\mathcal{P}U^{*}&({\bf k})\mathcal{P}^{\dagger} = \sigma_{x}\big[d_{0}\sigma_{0} + i(d_{x}\sigma_{x} - d_{y}\sigma_{y} + d_{z}\sigma_{z})\big]\sigma_{x}\\
& = d_{0}\sigma_{0} - i(-d_{x}\sigma_{x} - d_{y}\sigma_{y} + d_{z}\sigma_{z}) = U(-{\bf k}) \;,
\label{EQ:Symm}
\eal
\eea
since $d_{x,y}(-{\bf k}) = -d_{x,y}({\bf k})$ and $d_{z}(-{\bf k}) = d_{z}({\bf k})$ are odd and even functions of ${\bf k}$ respectively. On the other hand, it does not respect time-reversal ($\mathcal{T}$) and chiral ($\mathcal{C}$) symmetries in general. 

With only charge-conjugation symmetry ($\mathcal{P}^{2} = +1$), the system belongs to class D of topological classification and is characterized by a $\mathbb{Z}\times\mathbb{Z}$-topological index in two spatial dimensions \blue{\cite{Roy2017}}. Here $\mathbb{Z}\times\mathbb{Z}$ index refers to distinct integer invariant for each quasienergy gap (zero and $\pi$). \red{More importantly, such a symmetry also leads to the topological protection of counterpropagating edge states appearing in the system, as chiral edge states with positive and negative chirality are pinned around $k = 0$ and $k = \pi$ quasimomenta respectively, thus preventing their hybridization through continuous deformations. As an immediate consequence, such counterpropagating edge states can only be created or destroyed through a quasienergy gap closing and reopening process, which allows one to systematically characterize and control their number through the tuning of some system parameters as further detailed below.} 

In the following two sections, we show how various edge states with distinct features arise as the system parameter $J_{3}$ is varied. \red{In particular, we identify the occurrence of two types of band closings at several $J_3$ values, whose interplay leads to the emergence of counter-propagating edge states, characterizable by a combination of the two-terminal conductance \blue{\cite{ Fulga2016}} and dynamical winding number \blue{\cite{ Rudner2013}}.} Moreover, we find that at sufficiently large values of $J_3$, almost flat (dispersionless) edge states emerge in addition to the generation of new counter-propagating edge states. We further identify the emergence of chiral symmetry at (many) isolated points in the Brillouin zone, which explains the existence of these almost flat edge states. \red{It is expected that most of these features are also present in other charge-conjugation symmetry-protected Floquet topological insulators.}

\section{Counter-propagating Edge States}\label{sec_CP}
In this section, we show how the system introduced above may support an arbitrary number of counter-propagating edge states. To this end, we start by analytically solving the parameter values for which the quasienergy bands close, as well as the quasienergy dispersion and effective Hamiltonian near these band touching points. In particular, we observe that the two different types of band touching points that occur alternately as $J_3$ increases, leads to the generation of counter-propagating edge states. \red{Two-terminal conductance and dynamical winding number are used to establish the bulk-boundary correspondence.} Finally, symmetry-preserving disorder effects are studied in order to demonstrate the robustness of these edge states.

\subsection{Analysis of band touching points}\label{subsec_CP_bandclosing}
By diagonalizing the Floquet eigenvalue equation $U({\bf k})\mid\Psi_{\pm}({\bf k})\rangle = e^{-i\Omega_{\pm}({\bf k})}\mid\Psi_{\pm}({\bf k})\rangle$ and using Eq. (\blue{\ref{EQ:Floquet}}), the quasienergy expression can be obtained as, $\Omega_{\pm}({\bf k}) = \pm\cosinv(d_{0})$ where
\begin{equation}
d_{0} = \cos[\gamma_{1}]\cos[\gamma_{2}]\cos[\gamma_{3}] + \sin[\gamma_{1}]\sin[\gamma_{2}]\sin[\gamma_{3}]\;.
\end{equation}
 From the above expression it is seen that in the two dimensional Brillouin zone (BZ), the quasienergy bands close at either zero or $\pi$ quasienergy for arbitrary values of $J_{1}, J_{2}$ whenever $J_{3} = m\pi/3$, where $m$ is an integer. Under these parameter values, we may Taylor expand the Floquet operator near any BZ point $(k_{x_{0}},k_{y_{0}})$ such as $(k_{x}, k_{y}) = (k_{x_{0}} + \delta_{x}, k_{y_{0}} + \delta_{y})$ with $\delta_{x}(\delta_{y})$ being small deviations. In particular, we take the first order approximation in $\delta_{x}(\delta_{y})$ such that $\delta_{x}^{2}(\delta_{y}^{2}) \approx 0$ around $(k_{x_{0}}, k_{y_{0}}) = (0,0)$ to obtain
\bea
U(0 + \delta_{x}, 0 + \delta_{y}) = \pm\sigma_{0} -i( \pm J_{1}\delta_{x}\sigma_{x} \pm J_{2}\delta_{y}\sigma_{y})\;,
~~\label{EQ:Expansion00}
\eea
where upper (lower) sign refer to even (odd) value of $m$ (see Appendix \blue{\ref{App_bandclosing}}).

From Eq.~(\blue{\ref{EQ:Expansion00}}), we observe the following. First, gap closing around quasienergy zero ($\pi$) occurs at $(k_{x_{0}}, k_{y_{0}}) = (0,0)$ whenever $m$ is even (odd). Second, The effective Hamiltonian associated with Eq.~(\blue{\ref{EQ:Expansion}}) is $H_{eff} = \frac{J_{1}\delta_{x}\sigma_{x} + J_{2}\delta_{y}\sigma_{y}}{\sqrt{(J_{1}\delta_{x})^{2} + (J_{2}\delta_{y})^{2}}}$, which takes the form of Dirac Hamiltonian. The Dirac-like effective Hamiltonian shows that quasienergy dispersion is linear near the band touching point.

We further observe that for $m = 3m'$ with $m'$ being an integer, additional band touching points appear in the two dimensional BZ along with $(k_{x_{0}},k_{y_{0}}) = (0,0)$. These points are $(k_{x_{0}},k_{y_{0}}) = (0,\pi), (\pi,0)$ and $(\pi,\pi)$. The Floquet operator around these points can be expanded up to first order in $\delta_x$ and $\delta_y$ as,
\bea
\bal
U(0 + \delta_{x}, \pi + \delta_{y}) &= \pm\sigma_{0} -i(\pm J_{1}\delta_{x}\sigma_{x} \mp J_{2}\delta_{y}\sigma_{y})\;, \\
U(\pi + \delta_{x}, 0 + \delta_{y}) &= \pm\sigma_{0} -i (\mp J_{1}\delta_{x}\sigma_{x} \pm J_{2}\delta_{y}\sigma_{y})\;, \\
U(\pi + \delta_{x}, \pi + \delta_{y}) &= \pm\sigma_{0} -i(\mp J_{1}\delta_{x}\sigma_{x} \mp J_{2}\delta_{y},\sigma_{y})\;,
\label{EQ:Expansion}
\eal
\eea
where upper (lower) sign again refers to even (odd) values of $m = 3m'$. The same observations above, namely, alternate gap closing around zero and $\pi$ quasienergies and Dirac effective Hamiltonian are also made from Eq.~(\blue{\ref{EQ:Expansion}}).

The above analysis shows that there are two types of topological phase transitions (referred to as type-I and type-II for the sake of naming them) occurring in the system as $J_{3}$ parameter is varied. Type-I (type-II) topological phase transitions refer to those involving gap closing at a single point $(k_{x_{0}},k_{y_{0}}) = (0, 0)$ (four points $(k_{x_{0}},k_{y_{0}}) = (0, 0),(0,\pi),(\pi,0),(\pi,\pi)$) in the two dimensional BZ, which occur at $J_{3} = m\pi/3$ with $m\neq 3m'$ ($m=3m'$). It follows that as $J_3$ increases from zero, two type-I topological phase transitions first occur around $\pi$ and zero quasienergies at $J_3=\pi/3$ and $J_3=2\pi/3$ respectively, before first type-II topological phase transition occurs around $\pi$ quasienergy at $J_3=\pi$. Another two type-I topological phase transitions then occur around zero and $\pi$ quasienergies at $J_3=4\pi/3$ and $J_3=5\pi/3$ respectively, followed by a type-II topological phase transition at $J_3=2\pi$, now occurring around zero quasienergy. The same pattern described above then repeats itself as $J_3$ is varied further. As we demonstrate numerically below, the counter-propagating edge states emerge due to the alternate occurrences of type-I and type-II topological phase transitions at a given quasienergy.
\subsection{Topological characterization and bulk-boundary correspondence}\label{subsec_CP_TI}
In this section, we study the phase transitions induced by gap closing and reopening process as we vary the $J_{3}$ parameter. Topological phases are characterized by invariants to establish bulk-edge correspondence, which do not change as long as there is no gap closing. Our system supports both counter-propagating and chiral edge states which may or may not coexist. \red{In order to characterize the system completely, we compute the two-terminal conductance and dynamical winding number.} These indices respectively predict the total and net chirality \blue{\cite{Note1}} of edge states in a given band gap which helps us to determine the pairs of counter-propagating edge states.
\subsubsection{\red{Two-terminal conductance}}\label{subsubsec_CP_conductance}
We evaluate the system's two-terminal conductance \blue{\cite{ Fulga2016}} to characterize the total chirality of edge states in a quasienergy gap. To this end, we consider rectangular geometry such that the system has $N_{x}~(N_{y})$ number of unit cells in $x~(y)$ spatial direction. We apply external terminals in the form of absorbing boundary conditions in the $x$-direction such that the projector onto the absorbing terminal is given as,
\bea \nonumber
\bal
P = \begin{cases}
1 ~~~ &if~~ n_{y} \in \{1,N_{y}\}\;,\\
0 ~~~ &\text{otherwise}\;,
\end{cases}
\eal
\eea
where $n_{y}$ is the index to the unit cell in $y$-direction. The projector acts stroboscopically which is to say that the absorbing terminals only act at the beginning and end of each period. With these preliminaries, we define unitary scattering matrix for a given quasienergy gap $\epsilon \in \{0, \pi\}$ as,
\bea \nonumber
S^{\epsilon} = P\left[\mathbb{1} - e^{i\epsilon}\hat{U}(1-P^{T}P)\right]^{-1}e^{i\epsilon}\hat{U}P^{T} \;,
\eea
where $T$ denotes the matrix transpose and $\hat{U}$ being the Floquet operator under the boundary conditions defined above. The resulting scattering matrix is given as,
\bea
\bal
S^{\epsilon} = \left(\begin{array}{cc} r & t\\ t^{*} & r^{*} \end{array}\right) \;,
\eal
\eea
where $^{*}$ corresponds to the complex conjugation, $r$ and $t$ are the blocks of reflection and transmission amplitudes respectively. The two-terminal conductance is than given as a function of quasienergy as $G^{\epsilon} = \text{Trace}(tt^{*})$, where $\epsilon$ is taken in either zero or $\pi$ gap. It is worth mentioning that in realistic settings, an incoming state cannot be prepared at a given quasienergy $\epsilon$ value. Instead, an incoming state is prepared at a certain energy, and in that situation quantized conductance is obtained only after applying the so-called Floquet sum rule \blue{\cite{sumrule}}, which has been also demonstrated in Ref. \blue{\cite{Yap2017, Yap2018}}.

\subsubsection{Dynamical winding number}\label{subsubsec_CP_Winding}
Dynamical winding number characterizes the net chirality of edge states crossing zero and $\pi$ quasienergy gaps \blue{\cite{ Rudner2013}}. The idea is to determine the winding of edges states in quasienergy Brillouin zone without closing the $\epsilon$ gap, where $\epsilon \in \{0, \pi\}$. In order to calculate such an invariant, cyclic evolution is introduced by employing a modified time-evolution operator which is denoted by $\tilde{U}_{\epsilon}({\bf k},t)$ and given as,
\bea\nonumber
\bal
\tilde{U}_{\epsilon}({\bf k},t) =
\begin{cases}
U({\bf k},2t) ~~&if~~ 0 \le t  < T/2\\
e^{-iH^{\epsilon}_{eff}(2T-2t)} ~~&if~~ T/2 \le t  < T \;,
\end{cases}
\label{EQ:ReturnMap}
\eal
\eea
where $T = 1$ is the system's period, $H^{\epsilon}_{eff} = -\frac{i}{T}\log^{\epsilon}[U({\bf k},T)]$ is the effective Hamiltonian and $\epsilon$ is the branch cut of logarithm function which is taken to be the quasienergy gap under consideration. The operator during the second half of drive is a return map, which sends the modified time-evolution operator to identity at the end of one period, i.e., $\tilde{U}_{\epsilon}({\bf k}, t=0) = \tilde{U}_{\epsilon}({\bf k}, t=T) = \mathbf{1}$. Dynamical winding number is then given as \blue{\cite{ Rudner2013}},

\bea
\bal
W_{\epsilon} &= \frac{1}{8\pi^{2}}{\bf \int}dt~dk_{x}~dk_{y}\\
&\times Tr\Biggl( \tilde{U}^{-1}_{\epsilon}\partial_{t}\tilde{U}_{\epsilon}\Big[\tilde{U}^{-1}_{\epsilon}\partial_{k_{x}}\tilde{U}_{\epsilon}, \tilde{U}^{-1}_{\epsilon}\partial_{k_{y}}\tilde{U}_{\epsilon}\Big] \Biggl)\;,
\label{EQ:Winding}
\eal
\eea
here $W_{\epsilon}$ is the winding number in $\epsilon$ gap and square bracket represents the commutator.

The dynamical winding number is an integer equal to the net chirality of all chiral edge states crossing the $\epsilon$ gap. In particular, a pair of counter-propagating edge states has zero net chirality hence zero winding number. By contrast, a pair of co-propagating edge states carry two units of net chirality. Dynamical winding number predicts the net chirality of all the edge states crossing a particular quasienergy gap, so it may fail to capture the total number of edge states even when only one of the two (zero and $\pi$) quasienergy gaps possesses counter-propagating edge states.

\subsubsection{Bulk-boundary correspondance}\label{subsubsec_CP_results}
The two aforementioned topological invariants capture distinct features which are summarized as follows. \red{(i). Two-terminal conductance determines the total chirality \blue{\cite{Note1}} of edge states in a given quasienergy gap.} (ii). Dynamical winding number determines the net chirality of edge states in a gap such that two edge states with opposite chirality localized at the same edge give zero winding number. Following this observation, we define a quantity ($\mathcal{V}^{\epsilon}$) which determines the number of pairs of edge states carrying positive and negative chirality (i.e., pairs of counter-propagating edge states) around a quasienergy $\epsilon$ gap. This quantity is defined as,
\bea
\red{\mathcal{V}^{\epsilon} = \frac{G^{\epsilon} - \mid W_{\epsilon}\mid }{2}} \;,
\eea
where $\epsilon = \{0,\pi\}$ is the quasienergy gap, $G^{\epsilon}$ is the two-terminal conductance, and $W_{\epsilon}$ is the dynamical winding number.

\begin{figure}[H]
\centering
\includegraphics[width=0.95\linewidth, height=\linewidth, angle=270]{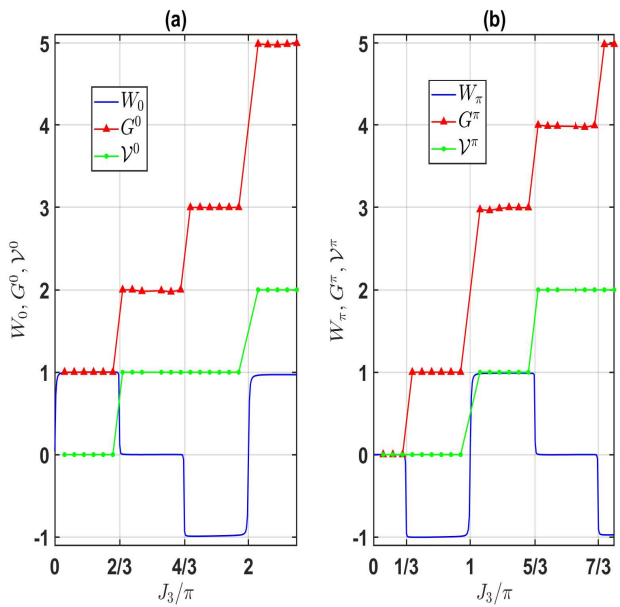}
\caption{Parameter values are $J_{1} = J_{2} = \pi/3,~ M = 1$. Dynamical winding number (blue), two-terminal conductance (red) and number of counter-propagating edge pairs (green) for (a) zero (b) $\pi$ quasienergy gap are shown as we vary $J_{3}$ parameter. Two-terminal conductance has been calculated in a rectangular sample with $N_{x} = 50,~ N_{y} = 170$ and absorbing boundary conditions are applied in the $y$-direction $n_{y} \in \{1, N_{y}\}$. Furthermore, for two-terminal conductance the value of $\epsilon$ is chosen in the gap such that $\epsilon = 0.0424\pi~\approx 0 $ for conductance $G^{0}$ in zero gap and $\epsilon = 0.9780\pi~\approx \pi $ for conductance $G^{\pi}$ in $\pi$ quasienergy gap.}
\label{Fig:Phase}
\end{figure}

The numerical calculations of two-terminal conductance $G^{\epsilon}$ (red curve), dynamical winding number $W_{\epsilon}$ (blue curve) and pairs of counter-propagating edge states $\mathcal{V}^{\epsilon}$ (green curve) are presented in Fig.~\blue{\ref{Fig:Phase}} for both zero and $\pi$ quasienergy gaps, as a function of $J_{3}$ parameter. It can be observed from Fig.~\blue{\ref{Fig:Phase}} that type-I and type-II topological phase transitions affect these invariants differently. Type-I topological phase transitions always decrease the dynamical winding number by a unit integer which results in an addition of one edge state with negative chirality at zero or $\pi$ quasienergy gap. \red{Simultaneously, two-terminal conductance increases by one unit. On the other hand, type-II topological phase transitions always add two edge states with positive chirality at zero or $\pi$ quasienergy gap which results in two units increase of both two-terminal conductance and dynamical winding number.} It is observed that two type-II topological phase transitions are separated by two type-I topological phase transitions and such a scheme lead to the creation of counter-propagation edge states in both zero and $\pi$ quasienergy gaps.

\begin{figure}[H]
\centering
\includegraphics[width=1.00\linewidth, height=\linewidth, angle=270]{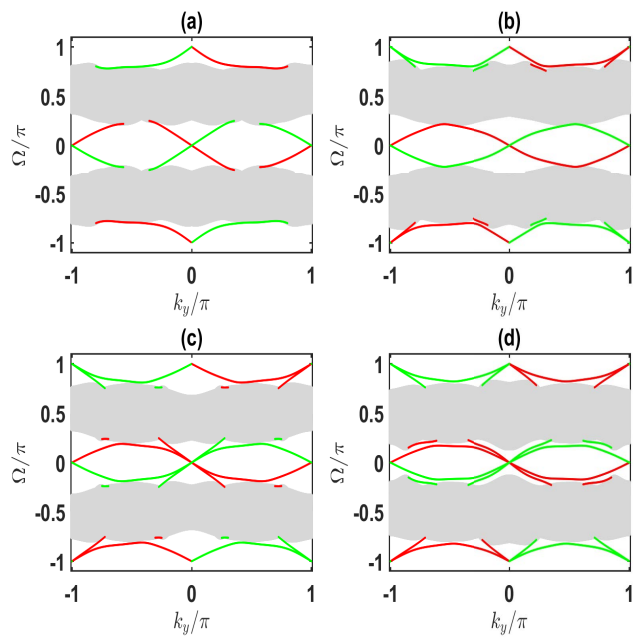}
\caption{Parameter values are $J_{1} = J_{2} = \pi/3, M = 1, N_{x} = 400$. States localized at the left (red) and right (green) edge for (a) $J_{3} = 0.8\pi$, (b) $J_{3} = 1.2\pi$,  (c) $J_{3} = 1.5\pi$, and (d) $J_{3} = 1.8\pi$ are shown. Two-terminal conductance, dynamical winding number and number of counter-propagating edge states around zero and $\pi$ quasienergy gap can be seen in Fig.~\blue{\ref{Fig:Phase}} for each distinct case.}
\label{Fig:CPES}
\end{figure}

In the following discussion, we will focus on the left edge of the system (red edge states in Fig. \blue{\ref{Fig:CPES}}) such that $n_{x} = 1$ under open (periodic) boundary conditions along $x~(y)$-direction respectively where $n_{x}$ is the unit cell index in $x$-direction. The first type-I phase transition occurs at $\pi$ quasienergy for $J_{3} = \pi/3$ which results in one edge state with negative chirality. From Fig.~\blue{\ref{Fig:Phase}(b)}, it can be observed that two-terminal conductance increases by unity while the dynamical winding number decreases by the same amount. The second type-I phase transition occurs at zero quasienergy for $J_{3} = 2\pi/3$, where conductance and dynamical winding number increases and decreases by one unit respectively due to the emergence of another edge state (now around zero quasienergy) with negative chirality. Together with the existing edge state of positive chirality around zero quasienergy, this results in the formation of a pair of counter-propagating edge states as can be observed in Fig.~\blue{\ref{Fig:CPES}(a)}. The counter-propagating pair gives rise to $\mathcal{V}^{0} = 1$ (green curve in Fig. \blue{\ref{Fig:Phase}(a)}). We also found that the emergence of such counter-propagating edge states does not depend on the direction in which open boundary conditions are applied, implying that they originate from strong topological effects (See Appendix \blue{\ref{App_sec:EdgeSpectrum}}). Moreover, the presence of charge-conjugation symmetry in two spatial dimensions also signal towards $\mathbb{Z}\times\mathbb{Z}$ topological classification \blue{\cite{Roy2017}}.

As we increase $J_{3}$ further, type-II phase transition occurs at $\pi$ quasienergy gap when $J_{3} = \pi$, which results in two additional edge states (around quasienergy $\pi$) with positive chirality. \red{All these three states around $\pi$ quasienergy gap (Fig.~\blue{\ref{Fig:CPES}(b)}), contribute towards $G^{\pi} =  3$.} The number of counter-propagating edge states and chiral edge states is given as $(\mathcal{V}^{\pi}, W_{\pi}) = (1, 1)$ in Fig.~\blue{\ref{Fig:Phase}(b)}. Further increase in $J_{3}$ parameter will induce two consecutive type-I phase transitions when $J_{3} = 4\pi/3$ and $J_{3} = 5\pi/3$ at zero and $\pi$ quasienergy gap respectively. Each phase transition will result in a decrease of dynamical winding number by a unit integer due to the addition of one edge state of negative chirality in the respective gap. Figure~\blue{\ref{Fig:CPES}(c)} shows the creation of one additional edge state with negative chirality around the zero quasienergy gap as a consequence of the phase transition at $J_{3} = 4\pi/3$. On the other hand, the phase transition at $J_{3} = 5\pi/3$ completes the pair of counter-propagating edge states around the $\pi$ quasienergy gap which are depicted in Fig.~\blue{\ref{Fig:CPES}(d)}. \red{The total chirality, number of chiral and counter-propagating edge states are respectively in full agreement with the calculated two-terminal conductance, dynamical winding number and the quantized quantity $\mathcal{V}^{\epsilon}$ presented in Fig.~\blue{\ref{Fig:Phase}}.}

Next phase transition occurs around the zero quasienergy gap when $J_{3} = 2\pi$, which completes the pair of counter-propagating edge states around this gap. Addition of two edge states with positive chirality increases the dynamical winding number form -1 to +1. The counter-propagating pairs and chiral edge states are given as $(\mathcal{V}^{0}, W_{0}) = (2, 1)$. As we keep increasing $J_{3}$ parameter for a large finite lattice, any number of counter-propagating edge states can be produced in a systematic way which are characterized by $\mathcal{V}^{\epsilon}$ around $\epsilon$ quasienergy gap.  For example, in the next section some explicit examples with three pairs of counter-propagating edge modes will be shown.    In addition to this, the presence or absence of chiral edge states along with the counter-propagating edge pairs is given by dynamical winding number $W_{\epsilon}$.  To our knowledge, this is the first detailed study of the systematic creation of arbitrary number of counter-propagating edge states.

It is worth mentioning that the helical edge states of the quantum spin Hall insulators \cite{Koenig2007, Roth2009} and the counter-propagating edge states discussed here are similar to each other up to spin degrees of freedom. Helical edge states also exhibit the localization of opposite chiralities at the same edge of the system. Moreover, the two-terminal transport study of quantum spin Hall insulators \cite{Roth2009, Dolcini2011} results in similar conductance measurements as we have observed for the counter-propagating edge states. These intriguing similarities among the behaviour of two systems, requires in depth study which is beyond the scope of this article.

In conclusion, we have seen that the systematic generation of arbitrary number of counter-propagating edge states around zero and $\pi$ quasienergy gaps is made possible through the presence of two types of topological phase transitions which occur alternately around a given quasienergy gap. \red{Moreover, the two-terminal conductance and dynamical winding number allow us to determine the number of pairs of counter-propagating edge states with certainty around both quasienergy zero and $\pi$ gaps through the quantity $\mathcal{V}^{\epsilon}$.}

\subsection{Disorder effects}
\label{subsec_Disorder}
In order to verify the robustness of the counter-propagating and chiral edge states in our system, we consider the presence of generic symmetry preserving disorders affecting the on-site mass term and the hopping term in both spatial directions during the third step of the quench. The hopping terms $J_{1}$ and $J_{2}$ during the first and second step of the quench are not subject to disorders for simplicity. Floquet operator for the disordered system is then given as $\hat{U} = e^{-i\hat{H}_{3}/3}e^{-i\hat{H}_{2}/3}e^{-i\hat{H}_{1}/3}$ with $\hat{H}_{1}, \hat{H}_{2}, \hat{H}_{3}$ being given as,
\bea
\bal\nonumber
\hat{H}_{1} &= \sum_{n_{x}}^{N_{x}}\sum_{n_{y}}^{N_{y}}\left[ -i\frac{3J_{1}}{2}\mid n_{x},n_{y}\rangle\langle n_{x} +1,n_{y}\mid - H.c\right]\sigma_{x}\;,\\
\hat{H}_{2} &= \sum_{n_{x}}^{N_{x}}\sum_{n_{y}}^{N_{y}}\left[~ -i\frac{3J_{2}}{2}\mid n_{x},n_{y}\rangle\langle n_{x},n_{y}+1\mid - H.c\right]\sigma_{y}\;,\\
\hat{H}_{3} &= \sum_{n_{x}}^{N_{x}}\sum_{n_{y}}^{N_{y}}3\Bigl[ \frac{(J_{3} + \lambda^{1}_{n_{x},n_{y}})}{2}\mid n_{x},n_{y}\rangle\langle n_{x} +1,n_{y}\mid\\
&~~~~~~~~~~ + \frac{(J_{3} + \lambda^{2}_{n_{x},n_{y}})}{2}\mid n_{x},n_{y}\rangle\langle n_{x},n_{y}+1\mid \\
&~~~~~~~~~~ + \frac{M(J_{3} + \lambda^{3}_{n_{x},n_{y}})}{2}\mid n_{x},n_{y}\rangle\langle n_{x},n_{y}\mid + H.c \Bigl]\sigma_{z}\;,
\label{EQ:real_Ham}
\eal
\eea
where $\lambda^{1}_{n_{x},n_{y}}, \lambda^{2}_{n_{x},n_{y}}$ and $\lambda^{3}_{n_{x},n_{y}}$ are random numbers drawn from same uniform distribution such that $(\lambda^{1}_{n_{x},n_{y}}, \lambda^{2}_{n_{x},n_{y}}, \lambda^{3}_{n_{x},n_{y}}) \in [-\lambda, \lambda]$, $\lambda$ denoting the strength of disorder.

\begin{figure}[H]
\centering
\includegraphics[width=0.85\linewidth, height=\linewidth, angle=270]{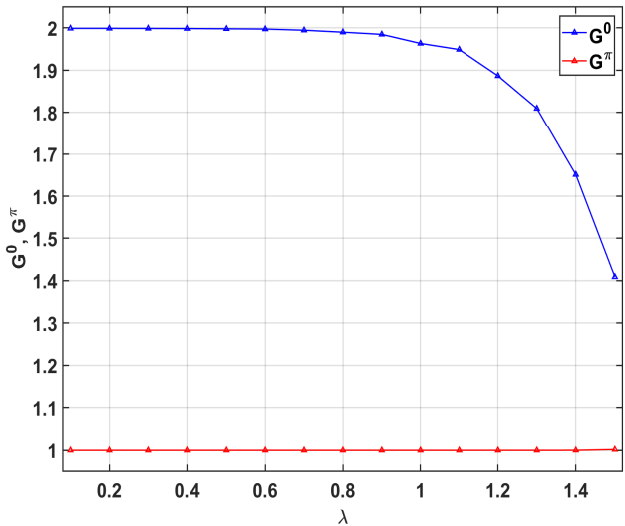}
\caption{Parameter values are $J_{1} = J_{2} = \pi/3,~M =1,~ N_{x} = N_{y} = 50$ and $J_{3} = 0.8\pi$. Two-terminal conductance $G^{\epsilon}$ around the zero (blue) and $\pi$ (red) quasienergy gap is averaged over $100$ disorder realizations. The value of $\epsilon$ is chosen in the gap such that $\epsilon = 0.0079\pi~\approx 0 $ for conductance $G^{0}$ around the zero quasienergy gap and $\epsilon = 0.9960\pi~\approx \pi $ for conductance $G^{\pi}$ around the $\pi$ quasienergy gap.}
\label{Fig:Disorder}
\end{figure}

We now consider the situation when we have a pair of counter-propagating edge states around the zero quasienergy gap and one chiral edge state around the $\pi$ quasienergy gap (See Fig. \blue{\ref{Fig:CPES}(a)}). We measure two-terminal conductance around both zero and $\pi$ quasienergy gap for the disordered lattice by using scattering matrix method \blue{\cite{ Fulga2016}}. Disorder effect on the counter-propagating edge states, which is reflected by the conductance around the zero quasienergy gap (indicated by the blue curve), can be observed in Fig.~\blue{\ref{Fig:Disorder}}. A plateau is clearly observed at small to moderate disorder strengths, indicating the robustness of such counter-propagating edge states. At larger disorder strengths, such as when $\lambda \gtrsim  1$, the two-terminal conductance starts to decrease exponentially. On the other hand, chiral edge state around the $\pi$ quasienergy gap is relatively more stable, which is reflected by the conductance around the $\pi$ quasienergy gap (indicated by the red curve) in Fig. \blue{\ref{Fig:Disorder}}. 

It is worth noticing that $\lambda = 1$ is a strong enough disorder strength for which $J_{3} + \lambda$ falls in another topological phase supporting a different number of edge states around both zero and $\pi$ quasienergy gaps (See Fig.~ \blue{\ref{Fig:Phase}}). As such, the deviation of $G^{0}$ from its quantized value at $\lambda \approx 0.8$ can be attributed to a disorder-induced topological phase transition rather than from the possible non-topological nature of counter-propagating edge states. Indeed, we have also verified (not shown in the figure) that around $\lambda \approx 0.8$, the bulk gap around quasienergy $0$ becomes very small, which further supports our previous argument.

In conclusion, we can argue that the pair of counter-propagating edge states and chiral edge states in our system are immune against weak to moderate symmetry preserving generic disorders. As the disorder strength increases, Disorder-induced topological phase transition takes place which leads to the deviation of $G^0$ and $G^\pi$ from their quantized values. In particular, a single chiral edge state, which exists at quasienergy $\pi$, is observed to be relatively more robust as compared with a pair of counter-propagating edge states existing at quasienergy $0$, as indicated by the fact that disorder-induced topological phase transition around quasienergy $\pi$ occurs at a larger disorder strength as comparated with that around quasienergy $0$. The relative difference in robustness between a pair of counter-propagating edge states and a single chiral edge state can be understood as follows. In the case of a pair of counter-propagating edge states, the boundaries of the system have two conducting channels and strong disorder can induce backscattering across these channels due to the absence of time reversal symmetry protection \blue{\cite{Wang2012}}. Additionally, localization effects can also contribute in decreasing the robustness of counter-propagating edge states. On the other hand, in the case of chiral edge states, the topology is mostly affected due to localization effects while backscattering is absent due to the one-way conducting channels.

\section{Almost Flat Edge States}\label{sec_flat_edge}
In this section, we show the emergence of chiral symmetry in the system at certain values of $J_{1}~ (J_{2})$ and large values of $J_{3}$ parameter, which leads to the emergence of almost flat edge states. We analytically find that how the chiral symmetry emerges in our system as we tune the $J_{3}$ parameter to large values and as a consequence almost flat (dispersionless) edge states appears at the systems' boundaries.

To demonstrate the emergence of chiral symmetry, we expand the Floquet operator at arbitrary points in the two dimensional BZ for certain parameter values. First, we choose the system parameter $J_{1} = \pi/2$ and expand the Floquet operator around BZ point $k_{x_{0}}$. We choose $k_{x_{0}} = \pi/2$, which is away from any band touching points. Next, we consider another BZ point parametrized by a deviation $\delta$ away from $k_{x_{0}} = \pi/2$ such that $k_{x} = k_{x_{0}} + \delta$. The resulting Floquet operator up to first order in $\delta$ is given as,
\bea
\bal
U(\delta, k_{y}) &= \Bigl[\sin(\gamma_{2})\sigma_{0} - i\cos(\gamma_{2})\sigma_{y}\Bigl]\sin(\gamma '_{3})\\
&+ \Bigl[-i\cos(\gamma_{2})\sigma_{x} + i\sin(\gamma_{2})\sigma_{z}\Bigl]\cos(\gamma '_{3}),
\label{Eq:FlatEdge}
\eal
\eea
where $\gamma_{2} = J_{2}\sin(k_{y})$, $\gamma '_{3} = J_{3}[M -\delta + \cos(k_{y}))$ and we fix $M = 1$ for simplicity. We observe from Eq. (\blue{\ref{Eq:FlatEdge}}) that for $\gamma '_{3} = m\pi$, the coefficient of $\sigma_{y}$ Pauli matrix will be zero i.e., $\sin(\gamma '_{3}) = 0$, where $m$ is an integer. The solution of all such points in the two dimensional BZ for fixed $\delta$ is given as $k_{y} = \cosinv[\frac{m\pi}{J_{3}} + \delta - 1]$ where $m \in [0,J_{3}/\pi]$ such that $-1 \le [\frac{m\pi}{J_{3}} + \delta - 1] \le 1$. We further observe that as we increase the value of $J_{3}$ parameter, the set of $m$ values becomes large for which the solution hold. The resulting Floquet operator is then given as,
\bea
\bal
U(\delta,k_{y}) = \mp i\cos(\gamma_{2})\sigma_{x} \pm i\sin(\gamma_{2})\sigma_{z}
\label{Eq:FlatEdge00}
\eal
\eea
where $\gamma_{2} = J_{2}\sin(k_{y}) = J_{2}\sin(\cosinv[\frac{m\pi}{J_{3}} + \delta - 1])$ and upper (lower) sign refers to even (odd) values of $m$. The unitary chiral symmetry operator is given as $\mathcal{C} = \sigma_{y}$ such that the Floquet operator obeys the chiral symmetry constraint $\mathcal{C}U(\delta,k_{y})\mathcal{C}^{\dagger} = U^{\dagger}(\delta, k_{y})$. We can repeat the same analysis for other points in the BZ such as $k_{x_{0}} = -\pi/2$ and observe the similar behaviour. It is important to mention that chiral symmetry emerges away from the band touching points in the two dimensional BZ. The existing edge states are expected to have almost flat dispersion because of this emergent chiral symmetry, namely, the system does have chiral symmetry at many isolated points in the BZ. The topological phase transitions will result in new chiral edge states near the band closing points as discussed in section \blue{\ref{sec_CP}}, irrespective of the presence or absence of chiral symmetry. The above analysis can be repeated by fixing $J_{2} = \pi/2$ instead of $J_{1}$ and the resulting Floquet operator is given as,
\bea
\bal\nonumber
U(k_{x}, \pi/2 + \delta) &= \Bigl[ \sin(\gamma_{1})\sigma_{0} + i\cos(\gamma_{1})\sigma_{x}\Bigl]\sin(\gamma '_{3})\\
& + \Bigl[ -i\cos(\gamma_{1})\sigma_{y} + i\sin(\gamma_{1})\sigma_{z}\Bigl]\cos(\gamma '_{3})\;.
\eal
\eea
where $\gamma_{1} = J_{1}\sin(k_{x})$, $\gamma_{3} = J_{3}(1 - \delta + \cos(k_{x}))$. Repeating the previous calculation results in the Floquet operator as,
\bea
\bal\nonumber
U(k_{x}, \delta) = \mp i\cos(\gamma_{1})\sigma_{y} \pm i\sin(\gamma_{1})\sigma_{x} \;,
\eal
\eea
and $\mathcal{C} = \sigma_{x}$ serves as the emergent chiral symmetry operator.

\begin{figure}[H]
\centering
\includegraphics[width=0.95\linewidth, height=\linewidth, angle=270]{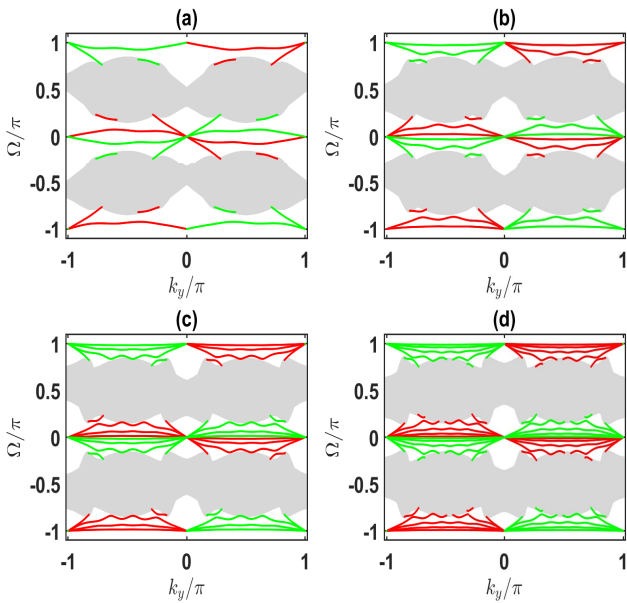}
\caption{Parameter values are $M = 1,~ J_{1} = \pi/2,~J_{2} = \pi/3$ and $N_{x} = 800$. Edge spectrum for (a) $J_{3} = 1.5\pi$, (b) $J_{3} = 2.5\pi$, (c) $J_{3} = 3.5\pi$, and (d) $J_{3} = 4.5\pi$ with open boundary conditions along $x$-direction and edge states localized at left (red) and right (green) ends are shown.}
\label{Fig:Flat_edge}
\end{figure}

In Fig.~\blue{\ref{Fig:Flat_edge}}, the full quasienergy spectra of our system at several $J_3$ values are shown under open (periodic) boundary conditions along $x~(y)$ direction. At small parameter values $J_{3} = 1.5\pi$, panel (a) shows that the system possess chiral and a pair of counter-propagating edge states with no flat (dispersionless) edge states. As we increase $J_{3} = 2.5\pi$, we observe that the existing pair of counter-propagating edge states becomes almost flat (Fig. \blue{\ref{Fig:Flat_edge}(b)}) with the emergence of a new pair of counter-propagating edge states. It is worth mentioning that the emergence of almost flat edge states depends on the direction of open (periodic) boundary conditions. This indicates that these almost flat edge states are a consequence of weak topological effects (See Appendix-\blue{\ref{App_sec:FlatEdge}}).

These almost flat edge states can be explained in terms of weak emergent chiral symmetry. The chiral symmetry is weak in the sense that Eq.~(\blue{\ref{Eq:FlatEdge00}}) holds only for a relatively small set of $m$ values.  As we increase $J_{3}$ further, the set of $m$ values for which Eq.~(\blue{\ref{Eq:FlatEdge00}}) is satisfied becomes bigger such that the corresponding points can be found throughout the two dimensional BZ, and a strong chiral symmetry emerges in the system which results in the observation of almost flat edge states. They, along with chiral edge states and pairs of counter-propagating edge states can be observed in Fig.~\blue{\ref{Fig:Flat_edge}(c-d)} around both quasienergy zero and $\pi$ gaps. Connecting the example here with the previous section, if we choose $J_1=\pi/3$ instead of $J_1=\pi/2$, we would have obtained four pairs of counter-propagating edge states for $J_{3} = 4.5\pi$. 

It is important to emphasize that while the almost flat edge states observed at large values of $J_{3}$ originate from the counter-propagating edge states, it does not contradict the topological nature of counter-propagating edge states elucidated earlier. In particular, these edge states are not completely flat and thus still possess positive and negative chirality around $k=0$ and $k=\pi$ (respectively) as expected for the formation of counter-propagating edge states. In principle, one needs to tune $J_{3}$ to infinity in order to completely flat them out, where the system consequenty undergoes an infinite number of topological phase transitions, since the flattening of edge states and the emergence of new counter-propagating pairs are the consequence of phase transitions controlled by the parameter $J_3$. \red{Finally, we note that the almost flat edge states observed above are difficult to probe via transport properties capturable within the scattering matrix formalism \cite{Fulga2016}. Developing new topological invariants for characterizing such almost flat edge states may thus be an interesting aspect which is beyond the scope of this paper and will be left for future studies.}

\section{summary and conclusion}\label{sec_sum}

In this paper, via a theoretically simple model system,  nonequilibrium topological systems with charge-conjugation symmetry are shown to have the capacity to accommodate an arbitrary number of chiral and counter-propagating edge states as certain system parameters are tuned. These edge states may or may not coexist around zero and $\pi$ gaps. \red{Furthermore, the topological characterization of these counter-propagating edge states has been made via the two-terminal conductance and dynamical winding number.} Moreover, it has been found that unlike previous studies, the counter-propagating edge states investigated here originate from strong topological effects (owing to charge-conjugation symmetry protection), because they do not depend on the direction in which the system is opened.

We have also demonstrated that both the counter-propagating and chiral edge states in a disordered lattice are immune against small to moderate symmetry preserving disorder. Under larger disorder strength, however, chiral edge states are found to be more robust as compared with a  pair of counter-propagating edge states. Finally, under certain parameter values, we have observed the emergent chiral symmetry in the two-dimensional BZ, which leads to almost flat/dispersionless edge states. Interestingly, counter-propagating, chiral, and these almost flat edge states can co-exist in some parameter regime.

In the future, it would be interesting to comprehensively analyse the parameter regime where counter-propagating, chiral and dispersionless edge states co-exist in terms of new topological invariants. Moreover, a more extensive study of the system in the presence of very strong disorder might also be an interesting aspect to pursue, where potentially new disorder induced Floquet topological features can be explored. Lastly, a study detailing some comparisons between counter-propagating edge states, which are unique to Floquet systems, with helical edge states commonly found in static systems is also worthwhile, as both have at least one property in common, i.e., localization at the same edge of the system with opposite chirality.

\acknowledgments
It is a pleasure to acknowledge helpful discussions with Han Hoe Yap and Linhu Li. R.W.B is supported by the Australian Research Council Centre of Excellence for Engineered Quantum Systems (EQUS, CE170100009). J. Gong acknowledges support from Singapore Ministry of Education Academic Research Fund Tier I (WBS No. R- 144-000-353-112) and by Singapore NRF Grant No. NRF- NRFI2017-04 (WBS No. R-144-000-378-281).

\appendix
\section{Analysis of Floquet operator and band touching points}\label{App_bandclosing}
Using Euler's formula, we obtain the Floquet operator in momentum space as, $U({\bf k}) = U_{3}({\bf k})U_{2}({\bf k})U_{1}({\bf k})$ where,
\bea
\bal
U_{1}({\bf k}) &= \cos(\gamma_{1})\sigma_{0} - i\sin(\gamma_{1})\sigma_{x}\;,\\
U_{2}({\bf k}) &= \cos(\gamma_{2})\sigma_{0} - i\sin(\gamma_{2})\sigma_{y}\;,\\
U_{3}({\bf k}) &= \cos(\gamma_{3})\sigma_{0} - i\sin(\gamma_{3})\sigma_{z}\;,
\label{Eq:App_StepOperators}
\eal
\eea
and $\gamma_{1} = J_{1}\sin(k_{x})$, $\gamma_{2} = J_{2}\sin(k_{y})$, $\gamma_{3} = J_{3}[M + \cos(k_{x}) + \cos(k_{y}) ]$ are the function of $k_x$ and $k_y$, $\sigma_{0}$ is $2\times 2$ identity operator and $\sigma_{x}, \sigma_{y}$ and $\sigma_{z}$ are the Pauli matrices in the sub-lattice degrees of freedom. The Floquet operator can be written as,
\bea
\bal
U({\bf k}) = d_{0}\sigma_{0} -i (d_{x}\sigma_{x} + d_{y}\sigma_{y} + d_{z}\sigma_{z}),
\eal
\eea
where $d_{0}, d_{x} , d_{y}$ and $d_{z}$ are all real and given as,
\begin{flushleft}
\bea
\bal
d_{0} &= \cos[\gamma_{1}]\cos[\gamma_{2}]\cos[\gamma_{3}] + \sin[\gamma_{1}]\sin[\gamma_{2}]\sin[\gamma_{3}]\;,\\
d_{x} &= \sin[\gamma_{1}]\cos[\gamma_{2}]\cos[\gamma_{3}] - \cos[\gamma_{1}]\sin[\gamma_{2}]\sin[\gamma_{3}]\;,\\
d_{y} &= \cos[\gamma_{1}]\sin[\gamma_{2}]\cos[\gamma_{3}] + \sin[\gamma_{1}]\cos[\gamma_{2}]\sin[\gamma_{3}]\;,\\
d_{z} &= \cos[\gamma_{1}]\cos[\gamma_{2}]\sin[\gamma_{3}] -\sin[\gamma_{1}]\sin[\gamma_{2}]\cos[\gamma_{3}]\;.
~~~~~~~\label{App_EQ:Ds}
\eal
\eea
\end{flushleft}
It can be observed that $d_{0}$ and $d_{z}$ are even functions of $\mathbf{k}=\left(k_x,k_y\right)$, while $d_{x}$ and $d_{y}$ are odd functions of $\mathbf{k}=\left(k_x,k_y\right)$. The quasienergy of the system is given as $\Omega_{\pm}({\bf k}) = \pm\cosinv(d_{0})$ and the effective Hamiltonian is given as $H_{eff} = |\Omega|[\frac{d_{x}\sigma_{x} + d_{y}\sigma_{y} + d_{z}\sigma_{z}}{\sqrt{d^{2}_{x}+ d^{2}_{y}+ d^{2}_{z}}}]$, where $|\Omega| = \cosinv[d_{0}]$.

In order to analyse the system, we Taylor expand the Floquet operator at certain points in Brillouin zone with fixed $M = 1$ and arbitrary values of other parameters. We choose a point $(k_{x},k_{y}) = (k_{x_{0}} + \delta_{x}, k_{y_{0}} + \delta_{y})$ such that $\delta_{x}(\delta_{y})$ are small deviations and $\delta^{2}_{x}(\delta^{2}_{y}) \approx 0$ and first order expansion is valid. First, we consider $(k_{x_{0}}, k_{y_{0}}) = (0, 0)$ such that up to first order approximation $\sin(k_{x_{0}} + \delta_{x}) = \delta_{x},~ \sin(k_{y_{0}} +  \delta_{y}) = \delta_{y}, ~\cos(k_{x_{0}} + \delta_{x}) = 1$ and $\cos(k_{y_{0}} + \delta_{y}) = 1$. The expansion of the Floquet operator results in,
\begin{flushleft}
\bea
\bal\nonumber
U(\delta_{x},&\delta_{y}) = \cos(3J_{3})\sigma_{0} - i\Big( \sin(3J_{3})\sigma_{z}\\
& +\big[J_{1}\delta_{x}\cos(3J_{3}) - J_{2}\delta_{y}\sin(3J_{3})\big]\sigma_{x}\\
& +\big[J_{2}\delta_{y}\cos(3J_{3}) + J_{1}\delta_{x}\sin(3J_{3})\big]\sigma_{y}\Big)\;.
~~~~\label{App_EQ:00}
\eal
\eea
\end{flushleft}
The quasienergies at BZ point $(k_{x_{0}}, k_{y_{0}}) = (0, 0)$ is given as $\Omega_{\pm}(0, 0) = \pm\cosinv[\cos(3J_{3})] = \pm3J_{3}$ which implies that both bands touch at zero ($\pi$) quasienergy when $J_{3} = m\pi/3$ where $m$ is an even (odd) integer. The Floquet operator near these band touching points is given as,
\bea
\bal
U(\delta_{x}, \delta_{y}) = \pm\sigma_{0} -i (\pm iJ_{1}\delta_{x}\sigma_{x} \pm iJ_{2}\delta_{y}\sigma_{y} )\;,
\label{Eq:App_00Dirac}
\eal
\eea
where upper (lower) sign indicates even (odd) values of $m$. Effective Hamiltonian can be then obtained as,
\bea
\bal
H_{eff} = \frac{J_{1}\delta_{x}\sigma_{x} + J_{2}\delta_{y}\sigma_{y}}{\sqrt{(J_{1}\delta_{x})^{2} + (J_{2}\delta_{y})^{2}}}\;,
\label{Eq:App_Effective}
\eal
\eea
which takes the form of a Dirac Hamiltonian. Hence our analysis shows that the band touching point at $(k_{x_{0}}, k_{y_{0}}) = (0, 0)$ with $J_{3} = m\pi/3$ has linear dispersion.

Secondly, we consider the BZ point $(k_{x_{0}}, k_{y_{0}}) = (\pi, \pi)$. We expand the Floquet operator at this BZ point up to first order in $\delta_{x} (\delta_{y})$ such that $\sin(k_{x_{0}}+\delta_{x}) = -\delta_{x},~ \sin(k_{y_{0}} + \delta_{y}) = -\delta_{y}, ~\cos(k_{x_{0}} + \delta_{x}) = -1$ and $\cos(k_{y_{0}} + \delta_{y}) = -1$, so that
\begin{flushleft}
\bea
\bal\nonumber
U(\delta_{x},&\delta_{y}) = \cos(J_{3})\sigma_{0} - i\Big(-\sin(J_{3})\sigma_{z}\\
& +\big[-J_{1}\delta_{x}\cos(J_{3}) - J_{2}\delta_{y}\sin(J_{3})\big]\sigma_{x}\\
& +\big[-J_{2}\delta_{y}\cos(J_{3}) + J_{1}\delta_{x}\sin(J_{3})\big]\sigma_{y} \Big)\;.
~~~~\label{Wq:App_PiPi}
\eal
\eea
\end{flushleft}
The quasienergies at $\delta_x=\delta_y=0$ are given as $\Omega_{\pm}(\pi,\pi) = \pm \cosinv[\cos(J_{3})] = \pm J_{3}$ and they are equal (modulo $2\pi$) for $J_{3} = m\pi$ with $m$ being an integer. Near these band touching points, the Floquet operator is given as
\bea
\bal
U(\delta_x, \delta_y) = \pm\sigma_{0} -i( \mp J_{1}\delta_{x}\sigma_{x} \mp J_{2}\delta_{y}\sigma_{y} )\;,
\eal
\eea
where upper (lower) sign corresponds to even (odd) values of $m$. The effective Hamiltonian will then have the form of a Dirac Hamiltonian similar to Eq.~(\blue{\ref{Eq:App_Effective}}), which indicates linear dispersion of quasienergy. BZ points $(k_{x_{0}}, k_{y_{0}}) = (0, \pi)$ and $(\pi, 0)$ also behave in the similar way. Hence at $J_{3} = m\pi$, the gap close at the four points in the BZ as discussed in main text.

\section{Edge spectrum under open boundaries in the $y$-direction}\label{App_sec:EdgeSpectrum}
Previous studies \blue{\cite{Lababidi2014,Zhou2014a,Ho2014,Zhou2014, Fulga2016}} show that the emergence of counter-propagating edge states depends on the direction in which the system is opened, which is a signature of weak topological effects \blue{\cite{Yoshimura2013,Yoshimura2014,Guo2014,Kobayashi2013}}. In this appendix, we will demonstrate that the counter-propagating edge states observed in the main text are in fact independent on the direction of the system's boundaries. To this end, we first note that in Fig.~\blue{\ref{Fig:CPES}} of the main text, we have shown the edge spectrum with open boundary conditions along $x$-direction. Here, we will instead focus on the case where open boundary conditions are applied along $y$-direction and compare our results to those of Fig.~\blue{\ref{Fig:CPES}}.

\begin{figure}[H]
\centering
\includegraphics[width=0.77\linewidth, height=\linewidth, angle=270]{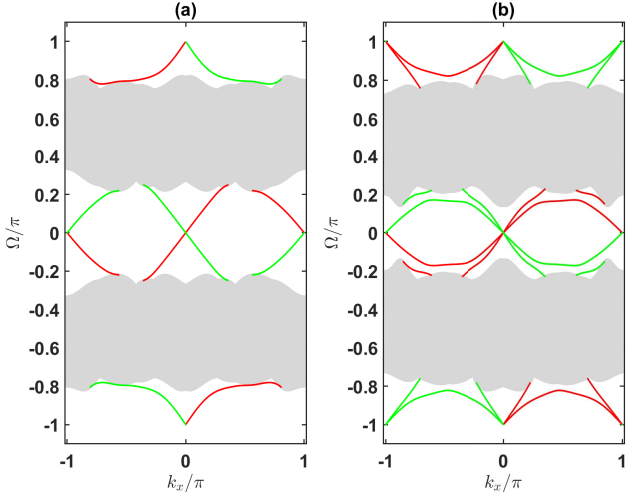}
\caption{Parameter values are $J_{1} = J_{2} = \pi/3, M = 1, N_{y} = 400$. Open (periodic) boundary conditions are applied in $y~(x)$-direction and states localized at left (red) and right (green) edge for (a) $J_{3} = 0.8\pi$, (b) $J_{3} = 1.8\pi$ have been shown.}
\label{Fig:App_CPES}
\end{figure}

Figure~\blue{\ref{Fig:App_CPES}} shows the full quasienergy spectrum of our system with open boundary conditions along $y$-direction. For $J_{3} = 0.8\pi$, one counter-propagating pair in zero gap and one chiral edge state in $\pi$ gap are shown in Fig.~\blue{\ref{Fig:App_CPES}(a)}. The phase corresponds to $(G^{0}, W_{0} )  = (2, 0)$ for zero gap and $(G^{\pi}, W_{\pi} )  = (1, -1)$ for the $\pi$ gap which can be observe in Fig. \blue{\ref{Fig:Phase}}. Increasing $J_{3}$ will induce type-I and type-II phase transitions which will result in the emergence of new chiral and counter-propagating edge states in the system. Figure~\blue{\ref{Fig:App_CPES}(b)} shows the parameter regime where our system exhibits two counter-propagating edge pairs around the $\pi$ quasienergy gap, as well as one pair of counter-propagating edge states and chiral edge states around the zero quasienergy gap for $J_{3} = 1.8\pi$. It corresponds to $(G^{0}, W_{0} )  = (3, -1)$ around the zero quasienergy gap and $(G^{\pi}, W_{\pi} )  = (4, 0)$ around the $\pi$ quasienergy gap  (Fig.~\blue{\ref{Fig:Phase}}). The above results thus show that the edge states in our system do not depend on the direction in which the system is opened.

\section{Robustness of dispersionless edge states}
\label{App_sec:FlatEdge}
In section \blue{\ref{sec_flat_edge}}, we have observed the dispersionless edge states under open (periodic) boundary conditions along $x~(y)$ spatial direction. It is important to mention that the dispersionless edge states in above mentioned case depend on the direction of open (periodic) boundary conditions. With the same parameter values, if we interchange the boundary conditions between two spatial directions, then only chiral and counter-propagating edge states are obtained at the systems' boundaries. Such a dependence on the boundary conditions indicates that the emergence of dispersionless edge states is a weak topological effect \blue{\cite{Yoshimura2013,Yoshimura2014,Guo2014,Kobayashi2013}}.

\begin{figure}[H]
\centering
\includegraphics[width=0.77\linewidth, height=\linewidth, angle=270]{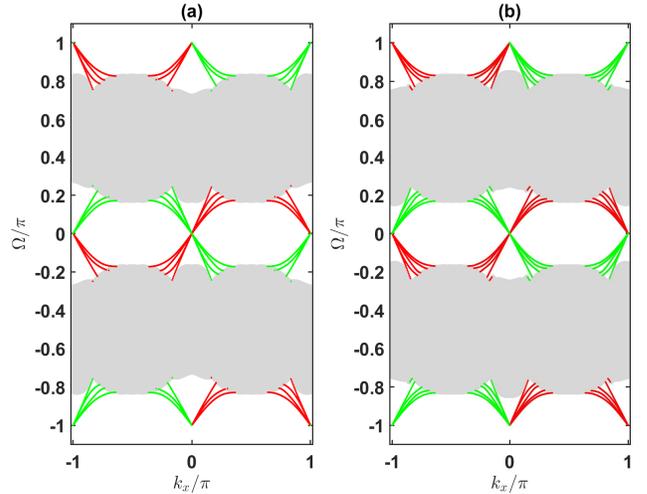}
\caption{Parameter values are $J_{1} = \pi/2,~J_{2} = \pi/3, M = 1, N_{y} = 800$. Open (periodic) boundary conditions are applied in $y~(x)$-direction and states localized at left (red) and right (green) edge for (a) $J_{3} = 3.5\pi$, (b) $J_{3} = 4.5\pi$ have been shown.}
\label{Fig:App_Flat}
\end{figure}

In order to demonstrate this weak topological effect, we choose $J_{1} = \pi/2$ and $J_{2} = \pi/3$ along with open (periodic) boundary conditions in $y~(x)$ direction. In Fig. \blue{\ref{Fig:App_Flat}}, we have shown the full quasienergy spectrum under open (periodic) boundary conditions in $y~(x)$ direction. Figure \blue{\ref{Fig:App_Flat}(a)} shows three pairs of counter-propagating edge states along with one chiral edge state in each quasienergy gap for $J_{3} = 3.5\pi$ and spectrum does not carry any dispersionless edge state. As we increase $J_{3}$ by $\pi$, another pair of counter-propagating edge states emerges in the system which is shown is Fig. \blue{\ref{Fig:App_Flat}(b)}. By comping Fig. \blue{\ref{Fig:Flat_edge}(c-d)} to Fig. \blue{\ref{Fig:App_Flat}(a-b)} respectively, it is concluded that these dispersionless edge states are a weak topological effect.

\twocolumngrid
\bibliographystyle{apsrev4-2}

\end{document}